\documentclass[
preprint,
 amsmath,amssymb,
 aps,
]{revtex4}

\usepackage{braket}
\usepackage{graphicx}
\usepackage{dcolumn}
\usepackage{bm}
\usepackage{color,float}
\newcommand{\simgt}{\lower.5ex\hbox{$\; \buildrel > \over \sim \;$}}
\newcommand{\simlt}{\lower.5ex\hbox{$\; \buildrel < \over \sim \;$}}

\begin{document}

\preprint{APS/123-QED}
\author{Daisuke Miki}
\author{Akira Matsumura}
\author{Kazuhiro Yamamoto}

\affiliation{
Department of Physics, Kyushu University, 744 Motooka, Nishi-ku, Fukuoka, Japan
}

\date{\today}

\title{Entanglement and decoherence of massive particles due to gravity}

\begin{abstract}
We analyze the dynamics of a gravity-induced entanglement for $N$ massive particles. Considering the linear configuration of these particles, we investigate the entanglement between a specific pair of particles under the influence
of the gravitational interaction between the massive particles. 
As the particle number increases, the specific particle pair decoheres more easily
due to the gravitational interaction with other particles.
The time scale of the gravity-induced decoherence is analytically determined.
We also discuss the entanglement dynamics of initially entangled particles, which exemplify the monogamy of the gravity-induced entanglement.
\end{abstract}
\maketitle

\section{\label{sec:level1}INTRODUCTION}

The unification of quantum mechanics with general relativity is one of the most fundamental problems in theoretical physics \cite{Feynmann}. 
The superstring theory is a promising candidate for theorizing quantum gravity (e.g., \cite{Ooguri}). However, many unresolved issues remain. 
One of the reasons for this is that there are almost no experimental studies on quantum gravity. 
Only a few experiments have been conducted to test quantum mechanics in a classical gravitational field. 
The COW experiment was the first to show that a neutron follows quantum mechanics 
in a uniform gravitational field \cite{COW}.
An experiment of a bouncing neutron in a uniform gravitational field has also been useful for investigating quantum mechanics in a gravitational field (Ref. \cite{VVN,Ichikawa},
cf.~\cite{Rohim}).
However, the experiments thus far have not answered the question of whether gravity 
follows quantum mechanics.

Recent advances in quantum sciences have opened the possibility of 
testing the quantum properties of gravity \cite{TabletopExperiments,Matsumoto1,Matsumoto2}. 
An interesting approach to test the superposition principle in a 
gravitational potential was proposed \cite{Bose17,MarlettoVedral17}; this study investigated whether a quantum entanglement is generated 
by gravity. 
This proposal is based on a theorem in quantum information theory that quantum entanglement cannot be generated by 
local operations and classical communication (LOCC) \cite{Horodecki09}. 
For example, consider a quantum system composed of two subsystems A and B. Two local observers, Alice and Bob, perform arbitrary quantum operations on subsystems A and B, respectively, and send classical information to each other.
This process, called LOCC,
cannot increase the quantum entanglement 
between the subsystems A and B.
Hence, if an operation generates a quantum entanglement, it is not LOCC.
Non-local classical operation might generate an entanglement.
However, we 
assume gravity based on a local theory; therefore, the production of an entanglement through gravity means the quantumness of gravity.

In Refs. \cite{Bose17,MarlettoVedral17}, an experimental test of a gravity-induced entanglement in a matter-wave interferometer, called a BMV experiment, was proposed. In the experimental setup, two massive particles with spin are initially in superposed states, and the gravitational interaction between these particles induces a quantum entanglement. The entanglement is then detected by measuring the spin correlations. For the feasible detection of an entanglement, one requires the superposition of a mesoscopic particle. In Ref. \cite{BoseMorley}, an experimental
setup for realizing such a superposition is presented; further, in Ref. \cite{Marshman}, the origin of generating  
the quantum entanglement is discussed.
These studies have in turn stimulated several studies on testing the quantum properties of gravity \cite{Belenchia,
Christodoulou,AnastopoulousHu,AnastopoulousHu2,Grossardt,Thomas,Krisnanda,NguyenBernards20}.
Nguyen and Bernards proposed a setup similar to that of the BMV experiment \cite{NguyenBernards20}. They assumed two separated masses, each of which was superposed in the direction perpendicular to their separation. This model could be easily analyzed because   
the gravitational interaction was simplified by the symmetry of the configuration. 

In the present study, we extend the model proposed by Nguyen and Bernards 
\cite{NguyenBernards20} to include $N$ massive particles (see Fig. 1) arranged in a linear configuration. This arrangement enables us to compute the quantum state of the total system explicitly. 
Then, we investigate the many-body effects of gravity on a quantum entanglement. 
Because gravity is locally unscreened (gravity is a long-range force), it might be interesting to examine how a quantum system coupled to other massive particles is influenced by gravitational interaction.
We show that the gravity-induced entanglement between a specific pair of particles is degraded by the decoherence due to gravitational interaction with other massive particles.
We also find that entanglement monogamy appears in the model 
by assuming an initially entangled state.
The features of decoherence and entanglement monogamy are demonstrated in the model of $N$ massive particles following the superposition principle.

This remainder of this paper is organized as follows. In Sec. 2, we introduce
the $N$-particle system on a straight line. Each particle
is assumed to be in a superposed position state in the
direction perpendicular to the straight line (see Fig. 1).
We present the Hamiltonian of the system, which describes the gravitational interaction between the particles written in a simple form. We also present the reduced density matrix of a specific pair
in the system, for which we evaluate the time evolution of 
the entanglement negativity. 
In Sec. 3, we extend the model in the previous section to the two-dimensional case.
In Sec. 4, we consider the case in which the initial state is 
an entangled state. This state demonstrates a monogamous behavior due to gravity. 
In Sec. 5, the summary and conclusions are presented.
In Appendix A, we describe the construction of the Hamiltonian. 
In Appendix B, a review of the derivation of the Hamiltonian Eq. (\ref{1-3}) is presented.
In Appendix C, we present a proof of
the negativity of the eigenvalues of the partial transposed density matrix in Eq. (\ref{3-11}).
In Appendix D, we show the density matrix of the initially entangled state.
In Appendix E, we explain the positive partial transpose (PPT) criterion and negativity.

\begin{figure}[b]
\includegraphics[width=9.0cm]{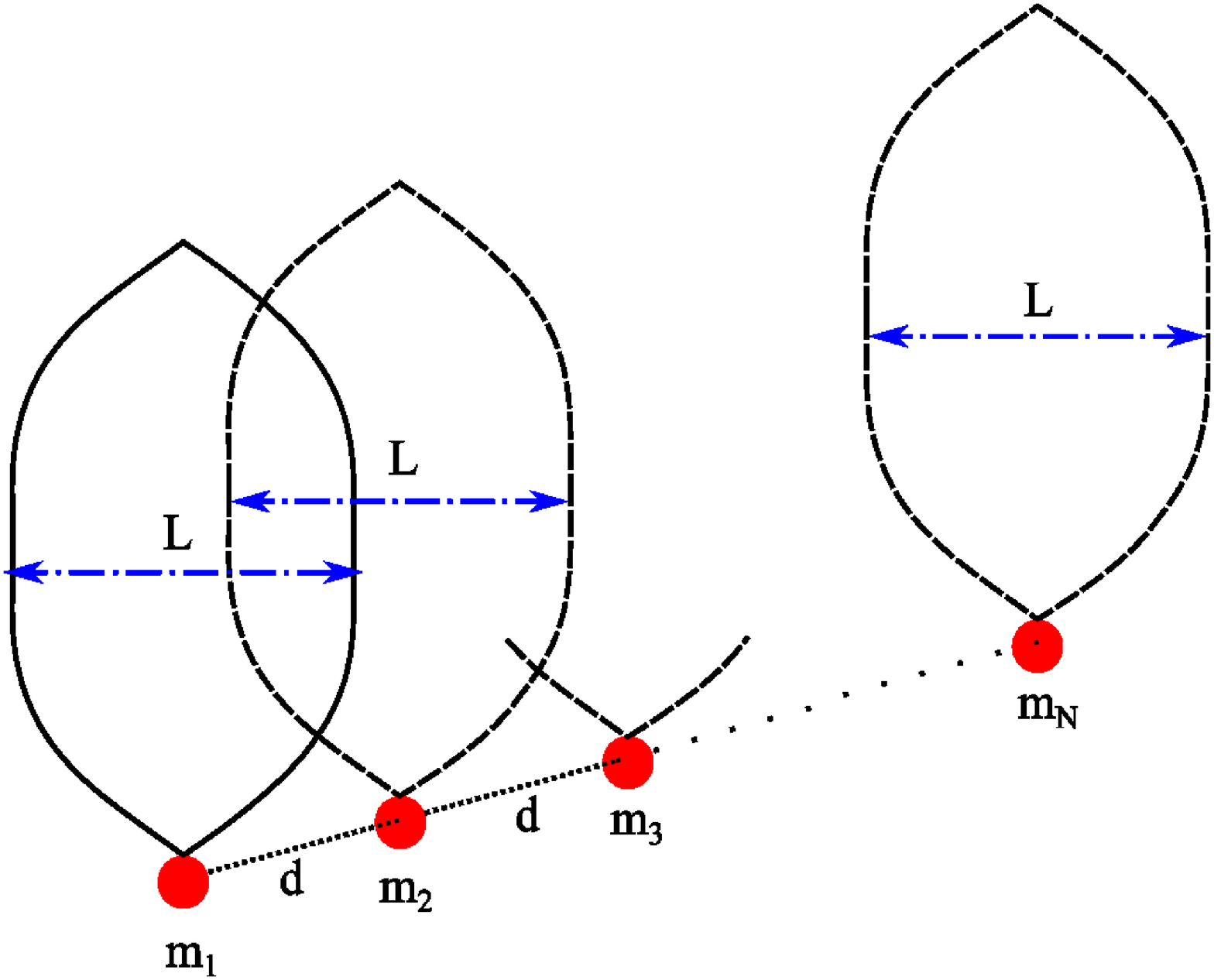}
\caption{Sketch of our model of the $N$-particle system.
Each particle is placed at a distance $d$ from its adjacent particles, and initially 
prepared in a superposition state with the position separated by distance $L$.
The direction of the separation between each particle $d$ is orthogonal to 
the direction of the separation of the particle with its superposed position $L$.
The $i$-th particle from the left ($1\leq i\leq N$) has mass $m_i$. The particles interact with one another through gravity.}
\label{fig:setup}
\end{figure}

\section{SYSTEM OF $N$ PARTICLES}

In this section, we introduce the system of $N$ massive particles to investigate 
the quantum nature of gravity. 
These particles are separate by a distance of $d$ from their immediate neighbors.
The $i$-th particle has mass $m_{i}$.
Each particle is initially prepared as the superposition of two spatially 
localized states separated by distance $L$ along the same direction. 
We aligned the $N$ particles so that the superposition was along the vertical direction. 
This model, which is depicted in Fig. \ref{fig:setup}, is an extension of the model in Ref.~\cite{NguyenBernards20}, which considered the $N=2$ case.

We use notations $\ket{\uparrow_i}$ and $\ket{\downarrow_i}$ to represent
the states of the $i$-th particle at the left and right paths, respectively. 
We consider the case where the initial 
state of the total system is
\begin{align}
\label{1-1}
\ket{\Psi(0)}
&=\ket{\psi_1(0)}\otimes\cdots\otimes\ket{\psi_N(0)}
\end{align}
where $\ket{\psi_{i}(0)}$ is the initial state of the $i$-th particle  
\begin{align}
\label{1-2}
\ket{\psi_{i}(0)}
&=\frac{1}{\sqrt{2}}(\ket{\uparrow_{i}}+\ket{\downarrow_{i}}).
\end{align}
The initial state evolves under the gravitational interaction. The corresponding Hamiltonian is
\begin{align}
    \label{1-3}
    H
    =\sum_{i<j}^{N}H_{ij}, 
    \quad H_{ij}
    =-\frac{\Delta_{ij}}{2}I_{1}\otimes\cdots\otimes\sigma_{z}^{(i)}\otimes\cdots\otimes\sigma_{z}^{(j)}\otimes\cdots\otimes I_{N}
\end{align}
where $H_{ij}$ (up to a constant) describes the Newtonian potential between the $i$- and $j$-th particles and $\Delta_{ij}$ for $i<j$ is
\begin{align}
    \label{1-5}
    \Delta_{ij}
    &=Gm_im_j \Bigl(\frac{1}{d(j-i)}-\frac{1}{\sqrt{\left(i-j\right)^2d^2+L^2}} \Bigr).
\end{align}
Because we assume that the wave packet of each particle does not spread, the kinetic term is neglected.
In Appendix A, we show that the familiar Hamiltonian from the Newtonian potential is described by the combination of $H_{ij}$ and another term that only contributes to a total phase, which we omit. 

The state of the total system at time $t$ is
$\ket{\Psi(t)}=e^{-iHt/\hbar}\ket{\Psi(0)}$.
The density matrix of the total system is obtained by explicit computation.
Here, we focus on the entanglement between the 1st and 2nd particles. 
Tracing over the 3rd to the $N$-th particles in the density operator $\rho(t)=\ket{\Psi(t)}\bra{\Psi(t)}$, we obtain 
the reduced density matrix of the 1st and 2nd particles as
\begin{widetext}
\begin{align}
\label{1-7}
\rho_{12}(t)
&=\text{Tr}_{3,\cdots,N}[\rho(t)]\notag\\
&=\frac{1}{4}
\left(
\begin{array}{cccc}
1&e^{\frac{i \Delta_{12}t}{\hbar}}\prod\limits_{i=3}^{N}\cos(\frac{\Delta_{2i}t}{\hbar})&e^{\frac{i \Delta_{12}t}{\hbar}}\prod\limits_{i=3}^{N}\cos(\frac{\Delta_{1i}t}{\hbar})&\prod\limits_{i=3}^{N}\cos(\frac{(\Delta_{1i}+\Delta_{2i})t}{\hbar})\\
e^{-\frac{i \Delta_{12}t}{\hbar}}\prod\limits_{i=3}^{N}\cos(\frac{\Delta_{2i}t}{\hbar})&1&\prod\limits_{i=3}^{N}\cos(\frac{(\Delta_{1i}-\Delta_{2i})t}{\hbar})&e^{-\frac{i \Delta_{12}t}{\hbar}}\prod\limits_{i=3}^{N}\cos(\frac{\Delta_{1i}t}{\hbar})\\
e^{-\frac{i \Delta_{12}t}{\hbar}}\prod\limits_{i=3}^{N}\cos(\frac{\Delta_{1i}t}{\hbar})&\prod\limits_{i=3}^{N}\cos(\frac{(\Delta_{1i}-\Delta_{2i})t}{\hbar})&1&e^{-\frac{i \Delta_{12}t}{\hbar}}\prod\limits_{i=3}^{N}\cos(\frac{\Delta_{2i}t}{\hbar})\\
\prod\limits_{i=3}^{N}\cos(\frac{(\Delta_{1i}+\Delta_{2i})t}{\hbar})&e^{\frac{i \Delta_{12}t}{\hbar}}\prod\limits_{i=3}^{N}\cos(\frac{\Delta_{1i}t}{\hbar})&e^{\frac{i \Delta_{12}t}{\hbar}}\prod\limits_{i=3}^{N}\cos(\frac{\Delta_{2i}t}{\hbar})&1
\end{array}
\right).
\end{align}
\end{widetext}
Here, the order of the basis is $\{\ket{\uparrow_1}\ket{\uparrow_2},~\ket{\uparrow_1}\ket{\downarrow_2},~\ket{\downarrow_1}\ket{\uparrow_2},~\ket{\downarrow_1}\ket{\downarrow_2}\}$.
In the following, we discuss the quantum entanglement due to Newtonian gravity for $N=3$ and $N>3$ after reviewing the case of $N=2$. 

\begin{figure}[b]
\centering
\includegraphics[width=8.0cm]{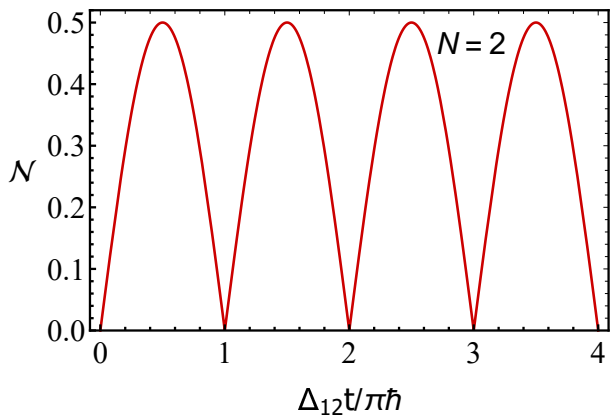}
\caption{
The negativity 
$\mathcal{N}$ computed from the partial transposed matrix Eq.~(\ref{2-6})
at $N=2$ as a function of the dimensionless time $\Delta_{12}t/\hbar$.
Because $\mathcal{N}$ is greater than or equal to zero, the state of two particles is entangled except when $\mathcal{N}=0$.}
\label{f1}
\end{figure}

\subsection{\label{sec:level2}{Two-particle system ($N=2$)}}

Here, we consider the system consisting of only two massive particles, which is 
the same as the model investigated in Ref.~\cite{NguyenBernards20}. 
For $N=2$, the initial state in Eq. \eqref{1-1} is
\begin{align}
\ket{\Psi(0)}
&=\ket{\psi_1 (0)}\otimes \ket{\psi_2 (0)}\notag
=\frac{1}{\sqrt{2}}(\ket{\uparrow_1}+\ket{\downarrow_1})\otimes\frac{1}{\sqrt{2}}(\ket{\uparrow_2}+\ket{\downarrow_2}),
\end{align}
and the Hamiltonian of the two particles is
\begin{align}
H
&=-\frac{\Delta_{12}}{2}\sigma^{(1)}_z\otimes \sigma^{(2)}_z\notag, 
\quad
\Delta_{12}=Gm_1m_2 \Bigl(\frac{1}{d}-\frac{1}{\sqrt{d^2+L^2}}\Bigr).
\end{align}
The density matrix of a given pure state is 
\begin{align}
\label{2-3}
\rho(t)
&=e^{-iHt/\hbar}\ket{\Psi(0)}\bra{\Psi(0)}e^{iHt/\hbar}\notag\\
&=\frac{1}{4}
\left(
\begin{array}{cccc}
1&e^{i\frac{\Delta_{12}}{\hbar}t}&e^{i\frac{\Delta_{12}}{\hbar}t}&1\\
e^{-i\frac{\Delta_{12}}{\hbar}t}&1&1&e^{-i\frac{\Delta_{12}}{\hbar}t}\\
e^{-i\frac{\Delta_{12}}{\hbar}t}&1&1&e^{-i\frac{\Delta_{12}}{\hbar}t}\\
1&e^{i\frac{\Delta_{12}}{\hbar}t}&e^{i\frac{\Delta_{12}}{\hbar}t}&1
\end{array}
\right).
\end{align}
We analyze the entanglement using the positive partial transpose (PPT) criterion \cite{Peres96}.
According to this criterion, the state is entangled if at least one of the eigenvalues of the partial transposed matrix of the density matrix is negative.
We now introduce the negativity defined as
\begin{align}
    \mathcal{N}
    =\sum_{\lambda_{i}<0}|\lambda_{i}|
    \label{2-5}
\end{align}
where the $\{\lambda_{i}\}$s are the eigenvalues of the partial transposed matrix.
The PPT criterion implies that the state is entangled if the negativity is positive, i.e., $\mathcal{N}>0$.
The eigenvalues of the partial transposed matrix are
\begin{align}
\label{2-4}
\lambda_{\pm}
&=\pm\frac{1}{2}
\sin \Bigl[\frac{\Delta_{12}}{\hbar}t \Bigr],
\quad
\lambda'_{\pm}
=\frac{1}{2}
\Bigl(
1\pm \cos \Bigl[\frac{\Delta_{12}}{\hbar}t \Bigr] 
\Bigr).
\end{align} 
$\lambda'_{\pm}$s are always positive, i.e., $\lambda'_{\pm}>0$. In contrast, either $\lambda_{+}$ or $\lambda_{-}$ is always negative or zero.
Therefore, the negativity is
\begin{align}
\label{2-6}
\mathcal{N}
&=\frac{1}{2}\left|\sin 
\Bigl[\frac{\Delta_{12}}{\hbar}t \Bigr]\right|.
\end{align}
Fig. \ref{f1} shows $\mathcal{N}$ as a function of the dimensionless time $\Delta_{12}t/\pi\hbar$.
The two particles periodically oscillate between the maximally entangled and non-entangled states. 
This is because the effects of the environment have not been considered \cite{NguyenBernards20}. 
\begin{figure}[b]
\includegraphics[width=8.0cm]{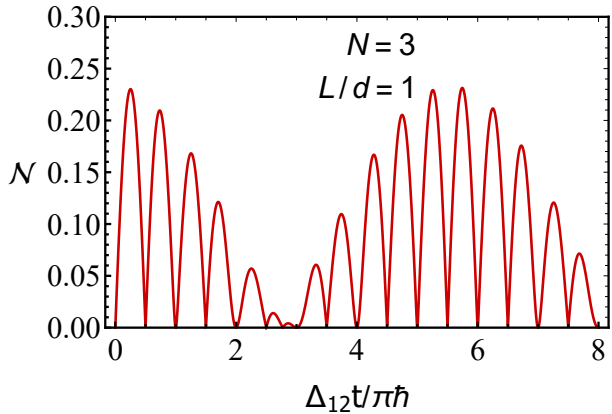}
\caption{
Behavior of 
$\mathcal{N}$ as a function of the dimensionless time $\Delta_{12}t/\pi\hbar$ for $L/d=1$ when the three particles have the same mass. In the case of $N=3$, $\mathcal{N}$ either takes positive values or is zero; hence, the 1st and 2nd particles are always entangled except when $\mathcal{N}=0$.}
\label{fig:three}
\end{figure}

\subsection{{Three-particle system ($N=3$)}}

Next, we consider the system with three massive particles $N=3$ and focus on the entanglement between two particles in this system. 
The initial state of the three particles is
\begin{align}
\label{3-1}
\ket{\Psi(0)}
&=\ket{\psi_1 (0)}\otimes\ket{\psi_2 (0)}\otimes\ket{\psi_3 (0)}\notag\\
&=\frac{1}{\sqrt{2}}(\ket{\uparrow_1}+\ket{\downarrow_1})\otimes\frac{1}{\sqrt{2}}(\ket{\uparrow_2}+\ket{\downarrow_2})\otimes\frac{1}{\sqrt{2}}(\ket{\uparrow_3}+\ket{\downarrow_3}).
\end{align}
The unitary evolution of this system is governed by $U(t)=\exp\left[-iHt/\hbar\right]$
with the Hamiltonian $H=H_{12}+H_{13}+H_{23}$.
By tracing over the 3rd particle, the reduced density matrix of the two particles ( density matrix Eq. \eqref{1-7} for $N=3$) is obtained as
\begin{widetext}
\begin{align}
\label{3-10}
\rho_{12}(t)
&=\frac{1}{4}
\left(
\begin{array}{cccc}
1
&e^{i\frac{\Delta_{12}}{\hbar}t}
\cos(\frac{\Delta_{23}}{\hbar}t)
&e^{i\frac{\Delta_{12}}{\hbar}t}
\cos(\frac{\Delta_{13}}{\hbar}t)
&\cos(\frac{\Delta_{13}+\Delta_{23}}{\hbar}t)\\
e^{-i\frac{\Delta_{12}}{\hbar}t}\cos(\frac{\Delta_{23}}{\hbar}t)&1&\cos(\frac{\Delta_{13}-\Delta_{23}}{\hbar}t)&e^{-i\frac{\Delta_{12}}{\hbar}t}\cos(\frac{\Delta_{13}}{\hbar}t)\\
e^{-i\frac{\Delta_{12}}{\hbar}t}\cos(\frac{\Delta_{13}}{\hbar}t)&\cos(\frac{\Delta_{13}-\Delta_{23}}{\hbar}t)&1&e^{-i\frac{\Delta_{12}}{\hbar}t}\cos(\frac{\Delta_{23}}{\hbar}t)\\
\cos(\frac{\Delta_{13}+\Delta_{23}}{\hbar}t)&e^{i\frac{\Delta_{12}}{\hbar}t}\cos(\frac{\Delta_{13}}{\hbar}t)&e^{i\frac{\Delta_{12}}{\hbar}t}\cos(\frac{\Delta_{23}}{\hbar}t)&1
\end{array}
\right).
\end{align}
\end{widetext}
We investigated the entanglement between the 1st and 2nd particles based on the PPT criterion. 
We can compute the eigenvalues of the partial transposed matrix in Eq. \eqref{3-10}. 
The four eigenvalues, Eq.~(\ref{lambdapm}), and Eq.~(\ref{lambdapm'}) are presented in Appendix C.
As shown in Appendix C, the negativity can be written as
\begin{align}
\label{3-11}
\mathcal{N}
=-\frac{1}{4}\Big(1&-\left|\cos(\frac{\Delta_{13}}{\hbar}t)\cos(\frac{\Delta_{23}}{\hbar}t)\right|\notag\\
&-\sqrt{1+\cos^2(\frac{\Delta_{13}}{\hbar}t)\cos^2(\frac{\Delta_{23}}{\hbar}t)-2\cos(2\frac{\Delta_{12}}{\hbar}t)\left|\cos(\frac{\Delta_{13}}{\hbar}t)\cos(\frac{\Delta_{23}}{\hbar}t)\right|}\Big).
\end{align}
Fig.~\ref{fig:three} shows $\mathcal{N}$ as a function of the dimensionless time $\Delta_{12}t/\pi\hbar$.
Here, we assume that the three particles have the same mass, and the distance between each particle is equal to the superposition distance, i.e., $L=d$. 
We find that 
the negativity $\mathcal{N}$ is positive (the 1st and 2nd particles are entangled) except at the zeros that appear periodically.
The maximum value of $\mathcal{N}$ varies and is smaller than $1/2$, unlike in the case of $N=2$.
These differences are caused by the gravitational interaction with the 3rd particle. 
The reduction of the entanglement can be understood as being due to the gravity-induced entanglement with the additional 3rd particle that plays the role of the environment. 

\subsection{\label{sec:citeref}$N$-particle system $(N>3)$}

\begin{figure}[H]
\includegraphics[width=7.0cm]{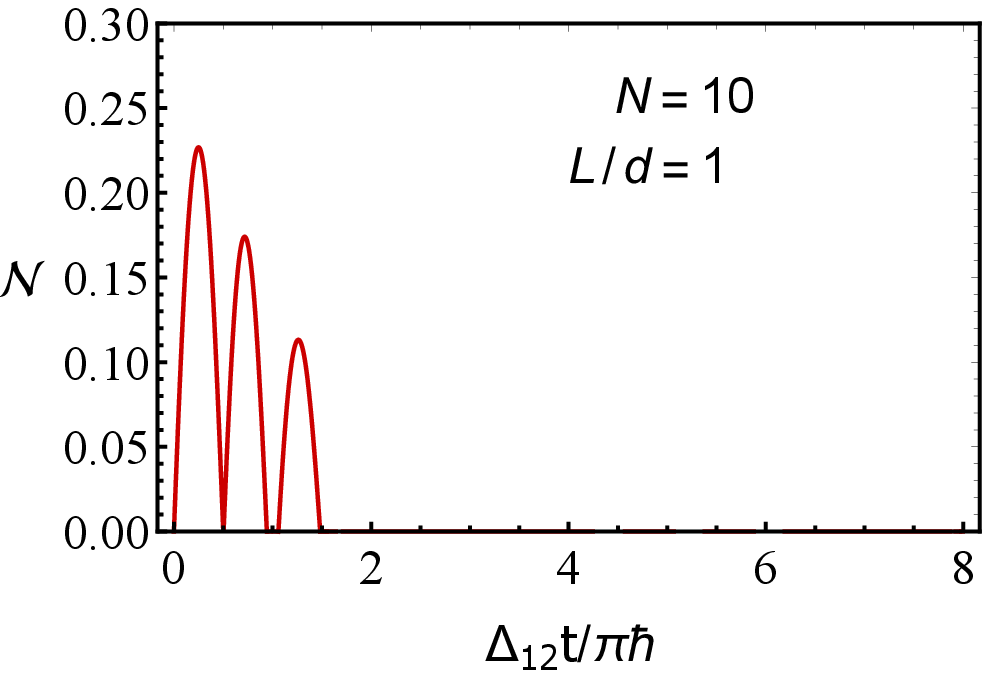}
\hspace{1cm}
\includegraphics[width=7.0cm]{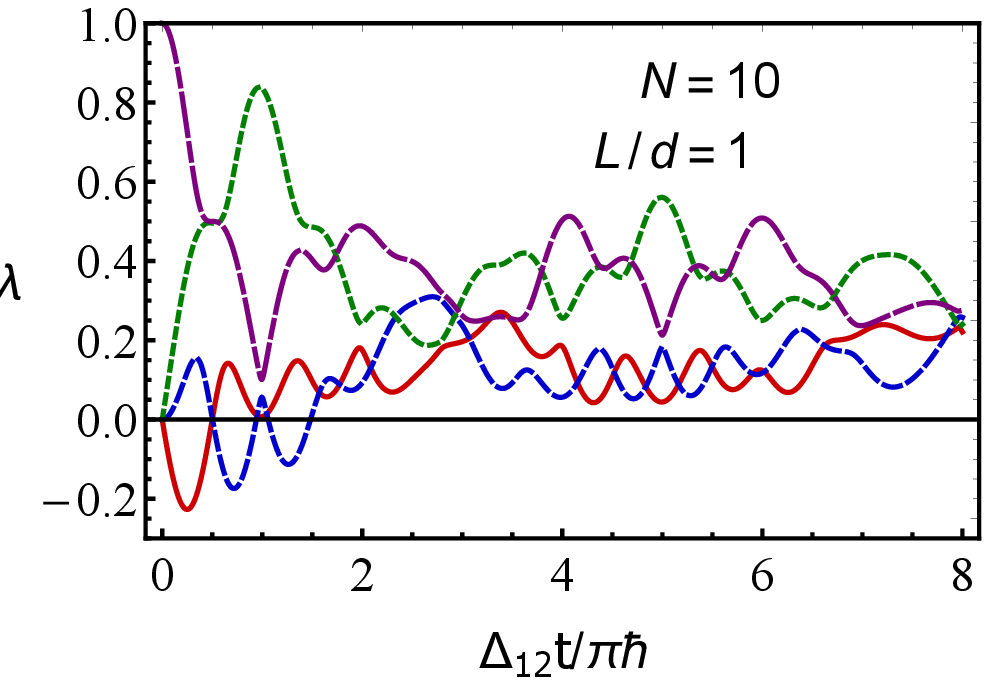}
\caption{
(left panel): Behavior of negativity, $\mathcal{N}$ 
as a function of the dimensionless time $\Delta_{12}t/\pi\hbar$.
Here, we assume ten particles ($N=10$) with the same mass and $L/d=1$. 
The negativity takes positive values for only a short period of $t\simlt 2\pi \hbar /\Delta_{12}$. 
(right panel): Behavior of the corresponding four eigenvalues (red solid line, blue dashed line, green dotted line, and purple dot-dash line). The negativity is positive if at least one of the eigenvalues is negative, and zero if all eigenvalues are positive.}
\label{f2A}
\end{figure}

In this subsection, we consider an $N$-particle system with more than 
three particles, i.e., $N>3$. 
We can compute the entanglement from the reduced density matrix in Eq.~(\ref{1-7}) with
respect to the 1st and 2nd particles.
The eigenvalues of the partial transposed matrix in Eq.~(\ref{1-7}) can be easily obtained. We find that two of the four eigenvalues can be negative. 
Here, we assume that all the particles have the same mass $m$. 
Figure \ref{f2A} demonstrates the evolution of the negativity and the four eigenvalues of the partial transposed matrix in Eq. \eqref{1-7} for $N=10$ and $L/d=1$. 
In contrast to the $N=3$ case, we see that the eigenvalues take negative values 
only for a short period after the initial time. 
One of the eigenvalues is negative at $t\simlt 2\pi \hbar /\Delta_{12}$, but both eigenvalues then
become positive. This means that the entanglement between the 1st and 2nd particles disappears at $t\simgt 2\pi \hbar /\Delta_{12}$. 
The gravitational interaction generates entanglement between the 1st and 2nd particles as well as the entanglement between these two particles and the other particles.
The result exemplifies the decoherence phenomenon due to gravity, although this decoherence is investigated in the framework of an open quantum system \cite{Schlosshauer1}.

\begin{figure}[htbp]
\includegraphics[width=7.0cm]{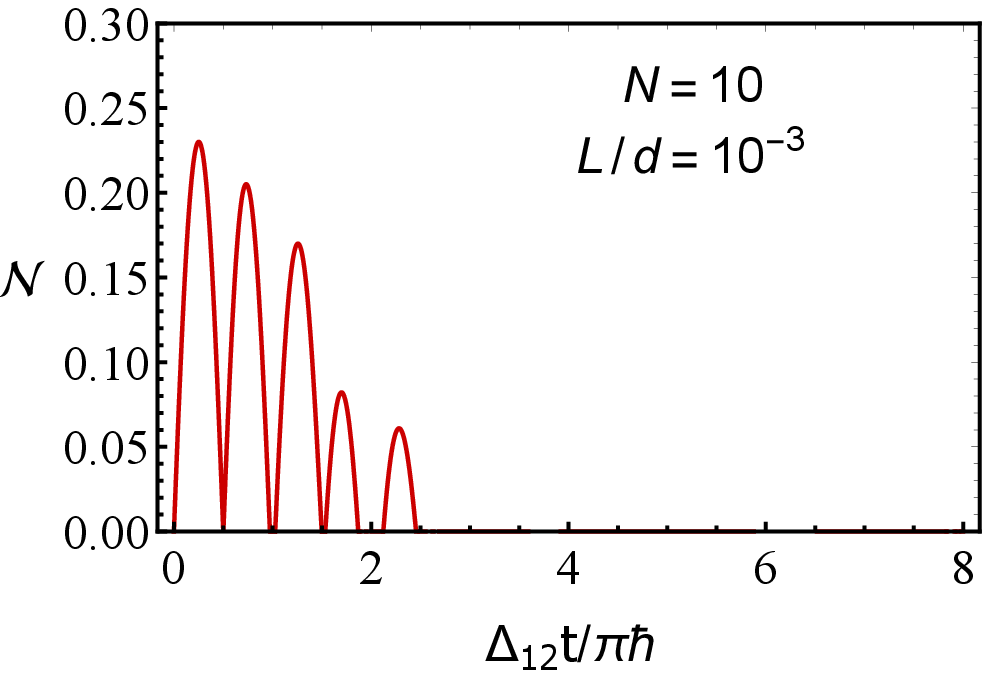}
\hspace{1cm}
\includegraphics[width=7.0cm]{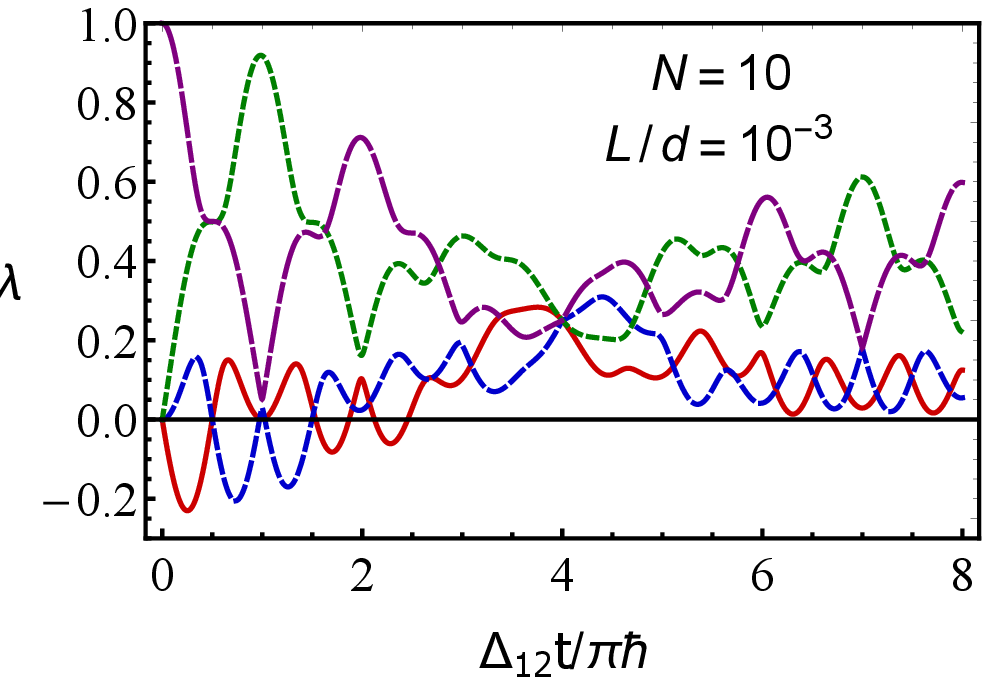}
\includegraphics[width=7.0cm]{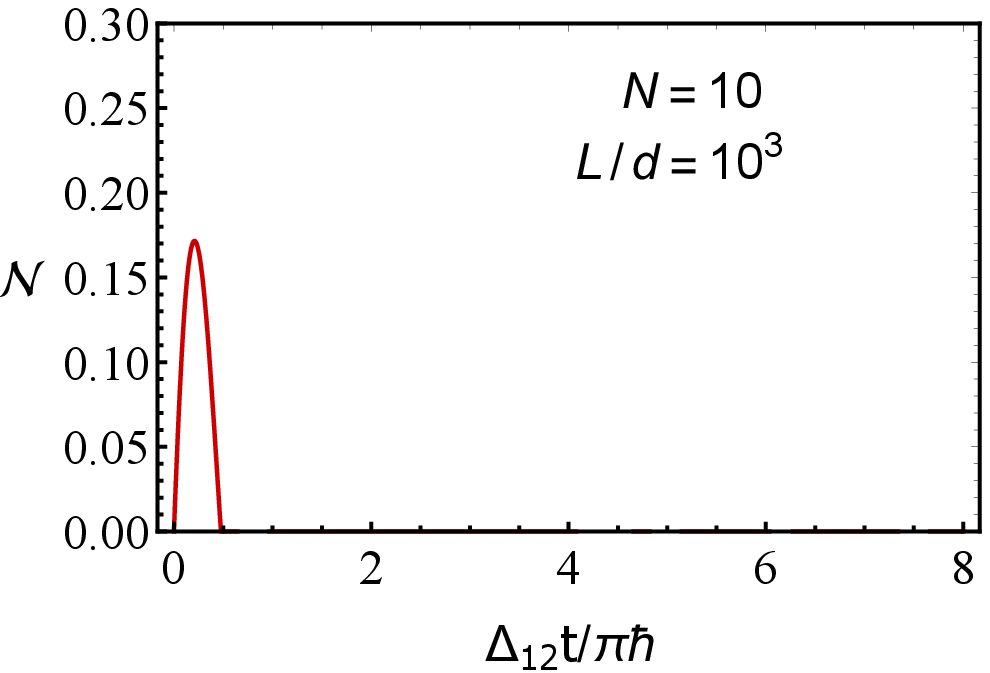}
\hspace{1cm}
\includegraphics[width=7.0cm]{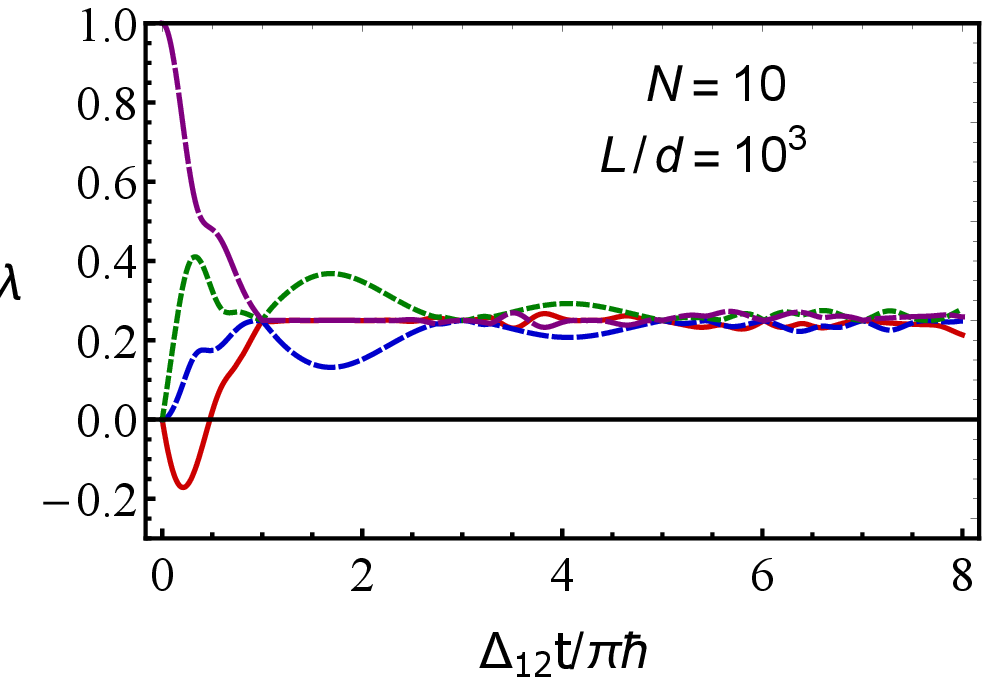}
\caption{
Same as Fig.\ref{f2A} but with $L/d=10^{-3}$ (upper left and right panels), and with $L/d=10^{3}$ (lower left and right panels).
\label{f2B}}
\end{figure}
The entanglement dynamics depend on 
ratio $L/d$ between the length scale of the superposition $L$ and the distance between the adjacent particles $d$.
Figure \ref{f2B} shows the time evolution of the negativity and eigenvalues of the partial transposed matrix Eq. \eqref{1-7}
at $L/d=10^{-3}$ (upper panels) and $L/d=10^{3}$ (lower panels).
The early entangled phase at $L/d=10^{3}$ lasts for a shorter time than at $L/d=1$ and $L/d=10^{-3}$. This is because the particles are close to one another when $L \gg d$, and the 1st and 2nd particles rapidly decohere because of the gravitational interaction with the other particles. By contrast, when $L\ll d$, the entanglement between the two particles 
is less likely to be affected by the other particles because the particles are far from one another. 
Hence, the entangled phase lasts longer than the other cases of $L/d=1$ and $L/d=10^3$. 

Let us examine the decoherence behavior analytically by taking the limit of the ratio $L/d$.
As the off-diagonal components of Eq.~(\ref{1-7}) characterize the coherence of the two particles, the decay time of these components determines that of the entanglement. 
From the inequality $\cos \theta\le e^{-\theta^2/2}$ for $0\le \theta\le\pi/2$, the absolute values of the off-diagonal components of Eq.~\eqref{1-7} satisfy
\begin{eqnarray}
\label{rho1234}
&&|\rho_{12}|=|\rho_{34}|=\prod_{i=3}^{N}\cos\left[\frac{\Delta_{2i}t}{\hbar}\right]
\le \exp\left[-\sum_{i=3}^{N} \frac{\Delta^2_{2i}t^2}{2\hbar^2} \right],
\\
\label{rho1324}
&&|\rho_{13}|=|\rho_{24}|=\prod_{i=3}^{\infty}\cos\left[\frac{\Delta_{1i}t}{\hbar}\right]
\le \exp\left[-\sum_{i=3}^{N}\frac{\Delta^2_{1i}t^2}{2\hbar^2}\right],
\\
\label{rho14}
&&|\rho_{14}|=\prod_{i=3}^{N}\cos\left[\frac{(\Delta_{1i}+\Delta_{2i})t}{\hbar}\right]
\le \exp\left[-\sum_{i=3}^{N}\frac{(\Delta_{1i}+\Delta_{2i})^2 t^2}{2\hbar^2}\right],
\\
\label{rho23}
&&|\rho_{23}|=\prod_{i=3}^{N}\cos\left[\frac{(\Delta_{1i}-\Delta_{2i})t}{\hbar}\right]
\le \exp\left[-\sum_{i=3}^{N}\frac{(\Delta_{1i}-\Delta_{2i})^2 t^2}{2\hbar^2}\right] , 
\end{eqnarray}
for 
$0 \leq (\Delta_{1i}\pm \Delta_{2i})t/\hbar \leq \pi/2$.
These inequalities enable us to examine the behavior of off-diagonal components. Under the condition $L\gg d$, the approximation
$\Delta_{ij}
\sim {Gm^2}/[{d(j-i)}]$
can be taken, and we can estimate 
the upper bounds of 
the absolute values of the off-diagonal components at $N\rightarrow \infty$ as
\begin{align}
\label{5-10}
|\rho_{12}|&=|\rho_{34}|
\le 
e^{-\scalebox{0.6}{$\displaystyle \sum_{i=3}^{\infty}\frac{1}{(i-2)^2}\left(\frac{Gm^2t}{d\hbar}\right)^2$}}
=
e^{-\scalebox{0.6}{$\displaystyle \zeta(2)\left(\frac{Gm^2t}{d\hbar}\right)^2$}},
\\
\label{5-7}
|\rho_{13}|&=|\rho_{24}|
\le 
e^{-\scalebox{0.6}{$\displaystyle \sum_{i=3}^{\infty}\frac{1}{(i-1)^2}\left(\frac{Gm^2t}{d\hbar}\right)^2$}}
=
e^{-\scalebox{0.6}{$\displaystyle (\zeta(2)-1)\left(\frac{Gm^2t}{d\hbar}\right)^2$}},
\\
\label{5-8}
|\rho_{14}|&
\le 
e^{-\scalebox{0.6}{$\displaystyle \sum_{i=3}^{\infty}
\left( \frac{1}{i-1}+\frac{1}{i-2} \right)^2 \left(\frac{Gm^2t}{d\hbar}\right)^2$}}
=
e^{-\scalebox{0.6}{$\displaystyle (2\zeta(2)+1)\left(\frac{Gm^2t}{d\hbar}\right)^2$}},
\\
\label{5-9}
|\rho_{23}|&
\le 
e^{-\scalebox{0.6}{$\displaystyle \sum_{i=3}^{\infty}
\left( \frac{1}{i-1}-\frac{1}{i-2} \right)^2 \left(\frac{Gm^2t}{d\hbar}\right)^2$}}
=
e^{-\scalebox{0.6}{$\displaystyle (2\zeta(2)-3)\left(\frac{Gm^2t}{d\hbar}\right)^2$}}, 
\end{align}
where $\zeta(n)$ is the zeta function. 
Therefore, we may write the decoherence time of our model when
$L\gg d$ as
\begin{eqnarray}
t_D\sim \frac{d\hbar}{Gm^2}\sim\frac{\hbar}{\Delta_{12}}.
\label{dtr}
\end{eqnarray}
In the lower left panel of Fig.~\ref{f2B}, which corresponds to  $L/d=10^{3}$, the negativity 
becomes zero when $\Delta_{12}t/\hbar\pi\simgt \mathcal{O}(1)$. This can be roughly explained by the
decoherence time Eq. (\ref{dtr}). Conversely, when $L\ll d$, the upper bounds are evaluated as follows: 
\begin{align}
\label{5-3}
|\rho_{12}|
&=|\rho_{34}|
\le 
e^{-\scalebox{0.6}{$\displaystyle \sum_{i=3}^{\infty}\frac{1}{8(i-2)^6}\left(\frac{Gm^2L^2t}{d^3\hbar}\right)^2$}}
=
e^{-\scalebox{0.6}{$\displaystyle \frac{\zeta(6)}{8}\left(\frac{Gm^2L^2t}{d^3\hbar}\right)^2$}},
\\
\label{5-4}
|\rho_{13}|&=|\rho_{24}|
\le 
e^{-\scalebox{0.6}{$\displaystyle \sum_{i=3}^{\infty} \frac{1}{8(i-1)^6}\left(\frac{Gm^2L^2t}{d^3\hbar}\right)^2$}}
=
e^{-\scalebox{0.6}{$\displaystyle \frac{\zeta(6)-1}{8}\left(\frac{Gm^2L^2t}{d^3\hbar}\right)^2 $}},
\\
\label{5-5}
|\rho_{14}|&
\le 
e^{-\scalebox{0.6}{$\displaystyle \sum_{i=3}^{\infty} \frac{1}{8} \left(\frac{1}{(i-1)^3}+\frac{1}{(i-2)^3} \right)^2 \left(\frac{Gm^2L^2t}{d^3\hbar}\right)^2$}}
=
e^{-\scalebox{0.6}{$\displaystyle\frac{1}{8} (2\zeta(6)-12\zeta(2)+19)\left(\frac{Gm^2L^2t}{d^3\hbar}\right)^2$}},
\\
\label{5-6}
|\rho_{23}|&
\le e^{-\scalebox{0.6}{$\displaystyle \sum_{i=3}^{\infty} \frac{1}{8} \left(\frac{1}{(i-1)^3}-\frac{1}{(i-2)^3} \right)^2 \left(\frac{Gm^2L^2t}{d^3\hbar}\right)^2$}}
=e^{-\scalebox{0.6}{$\displaystyle \frac{1}{8} (2\zeta(6)+12\zeta(2)-21)\left(\frac{Gm^2L^2t}{d^3\hbar}\right)^2$}}, 
\end{align}
where we used the approximation $\Delta_{ij}\sim {Gm^2L^2}/[2d(j-i)]^3$ and considered the limit where $N$ goes to infinity. 
Therefore, the decoherence time is approximately estimated as 
\begin{eqnarray}
t_D\sim \frac{d^3\hbar}{Gm^2L^2}\sim\frac{\hbar}{2\Delta_{12}}.
\label{eq:tD2}
\end{eqnarray}
In the left panel of Fig.~\ref{f2B}, which corresponds to $L/d=10^{-3}$, the negativity assumes a value of zero when $\Delta_{12}t/\hbar\pi\simgt\mathcal{O}(1)$.
This time scale roughly corresponds to the decay time \eqref{eq:tD2} of the off-diagonal components.

Thus, the decoherence time can be approximately evaluated using the decay rate of the off-diagonal components in the reduced density matrix. 
The above-mentioned results indicate that the decoherence time of our model does not strongly depend on the ratio of 
$L/d$, provided the number of particles is sufficiently large. 
This might be due to the special characteristics of the one-dimensional configuration of our model.

The state of a system with a finite number of degrees of freedom must evolve recursively.
However, the period is longer than the decoherence time due to the environment in a realistic case.

\section{Two-dimensional case}
\begin{figure}[htbp]
\includegraphics[width=7.0cm]{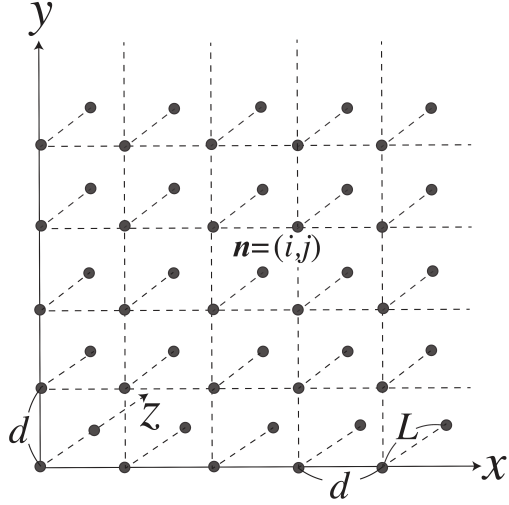}
\centering
\caption{
Alignment of particles in two dimensions.
Particles with mass $m$ are placed at lattice points with coordinates 
$(x,y)=d(i,j)$ and integers $i$ and $j$, where $d$ is the separation distance between the neighboring lattice points.
We assume that the total number of particles is $N\times N$ and that each particle is
specified by $\bm n=(i,j)$. 
This figure shows that each particle is in a superposition of localized states along the z-axis
separated by a distance $L$.
}
\label{2dim}
\end{figure}

The one-dimensional system discussed in the previous section can be generalized to a two-dimensional system in a simple way. Here, we consider a system consisting of $N\times N$ particles aligned on the $(x,y)$ plane, as illustrated in Figure \ref{2dim}.
The position of each particle on the $(x,y)$ coordinates is specified by $(x,y)= d\bm n=d(i,j)$ with integers $0\leq i \leq N$ and $0\leq j \leq N$. 
Each particle is prepared to be in a spatially localized superposition state 
separated by a distance $L$ along the direction of the z-axis.
The Hamiltonian of the system is
\begin{align}
H&=\frac{1}{2}
\sum_{\bm n,\bm n', (\bm n\neq \bm n')}H_{\bm n,\bm n'} ,
\end{align}
where $H_{\bm n,\bm n'}$ is defined by
\begin{align}
H_{\bm n,\bm n'}
&=-\frac{\Delta_{\bm n\bm n'}}{2} I\otimes\cdots\otimes I\otimes\sigma_{z}^{\bm n}\otimes I\otimes\cdots\otimes I\otimes\sigma_{z}^{\bm n'}\otimes I\otimes\cdots\otimes I \\
\Delta_{\bm n\bm n'}
&=Gm^2\left(\frac{1}{d \sqrt{|\bm n-\bm n'|^2}}-\frac{1}{\sqrt{d^2|\bm n-\bm n'|^2+L^2}}\right).
\end{align}

Here, we focus on the entanglement between the two particles 
at the points
$\bm n_0=(0,0)$ and 
$\bm n_1=(1,0)$. 
Similar to the one-dimensional case, we determine the reduced density matrix of these two particles as
\begin{align}
\rho_{\bm n_0\bm n_1}(t)
=\frac{1}{4}
\left(
\begin{array}{cccc}
1&\rho_{\bm n_0\bm n_1}^{12}&\rho_{\bm n_0\bm n_1}^{13}&\rho_{\bm n_0\bm n_1}^{14}\\
(\rho_{\bm n_0\bm n_1}^{12})^{*}&1&\rho_{\bm n_0\bm n_1}^{23}&\rho_{\bm n_0\bm n_1}^{24}\\
(\rho_{\bm n_0\bm n_1}^{13})^{*}&(\rho_{\bm n_0\bm n_1}^{23})^{*}&1&\rho_{\bm n_0\bm n_1}^{34}\\
(\rho_{\bm n_0\bm n_1}^{14})^{*}&(\rho_{\bm n_0\bm n_1}^{23})^{*}&(\rho_{\bm n_0\bm n_1}^{34})^{*}&1
\end{array}
\right),
\end{align}
where ${}^{*}$ represents the complex conjugate.
The components can be defined as
\begin{align}
\rho_{\bm n_0\bm n_1}^{12}
&=e^{\frac{i t\Delta_{\bm n_0\bm n_1 }}{\hbar}}\prod\limits_{\bm n}\cos(\frac{\Delta_{\bm n_1\bm n}t}{\hbar}),\\
\rho_{\bm n_0\bm n_1}^{13}
&=e^{\frac{i t\Delta_{\bm n_0\bm n_1}}{\hbar}}\prod\limits_{\bm n}\cos(\frac{\Delta_{\bm n_0\bm n}t}{\hbar}),\\
\rho_{\bm n_0\bm n_1}^{14}
&=\prod\limits_{\bm n}\cos(\frac{(\Delta_{\bm n_0\bm n} +\Delta_{\bm n_1\bm n})t}{\hbar}),\\
\rho_{\bm n_0\bm n_1}^{23}
&=\prod\limits_{\bm n}\cos(\frac{(\Delta_{\bm n_0\bm n}-\Delta_{\bm n_1\bm n})t}{\hbar}),\\
\rho_{\bm n_0\bm n_1}^{24}
&=e^{-\frac{it \Delta_{\bm n_0\bm n}}{\hbar}}\prod\limits_{\bm n}\cos(\frac{\Delta_{\bm n_0\bm n}t}{\hbar}),\\
\rho_{\bm n_0\bm n_1}^{34}
&=e^{-\frac{i t\Delta_{\bm n_0\bm n_1}}{\hbar}}\prod\limits_{\bm n}\cos(\frac{\Delta_{\bm n_1\bm n}t}{\hbar})
\end{align}
where $\prod_{\bm n}$ implies $\prod_{\bm n \neq \bm n_0,\bm n_1}$.

\begin{figure}[H]
\includegraphics[width=7.0cm]{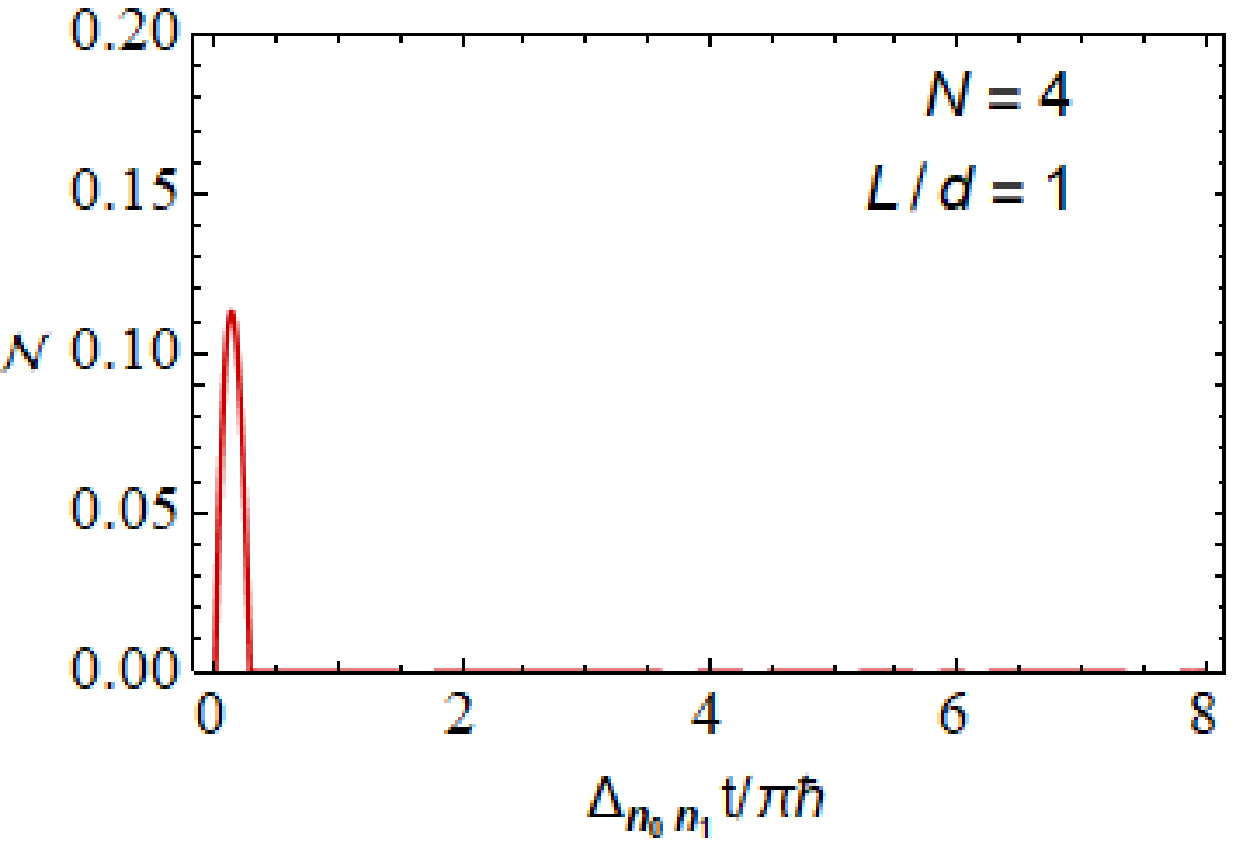}
\hspace{1cm}
\includegraphics[width=7.0cm]{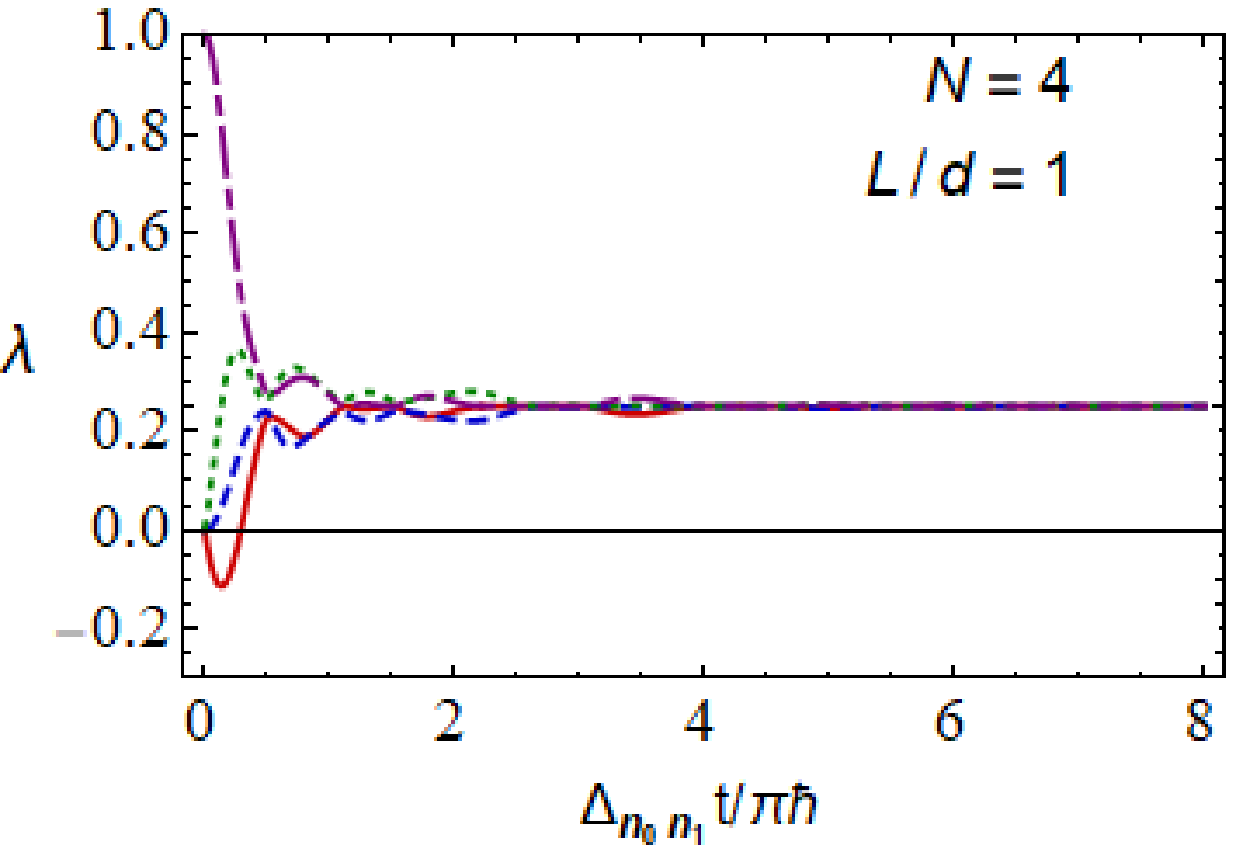}
\caption{
Same as Fig.4 but for the two-dimensional case with $N=4$.
 }
\label{4par2dim}
\end{figure}

Fig. \ref{4par2dim} depicts the result of the two-dimensional case 
for 
$N=4$ and 
$d=L$.
The entanglement disappears rapidly and a significant decoherence effect is observed in comparison with that in the one-dimensional case.
This is because the number of particles around the two particles
$\bm n_0$ and $\bm n_1$ increases, and the influence of the entanglement with the other particles becomes significant.

However, it is challenging to extend the above analysis to models with three-dimensional configurations, which we intend to explore in our future work.

\section{MONOGAMY OF INITIALLY ENTANGLED STATE}

The results presented in the previous sections demonstrate that the entanglement among a specific subsystem 
and the other systems causes quantum decoherence due to gravitational interaction. 
This feature can also be understood as an effect of monogamy in quantum systems. 
Here, we focus on monogamy due to gravity 
by considering the system prepared in an initial state, wherein the 
specific subsystem is 
entangled with other systems.

Specifically, in this section, we consider a one-dimensional system consisting of three particles prepared in their initial states, wherein the 2nd and 3rd particles are initially entangled, but are not entangled with the 1st particle. Therefore, we consider the initial states 
\begin{align}
\label{6-1}
\ket{\Psi(0)}
&=\frac{1}{\sqrt{2}}(\ket{\uparrow_1}+\ket{\downarrow_1})\otimes (a\ket{\uparrow_2}\ket{\downarrow_3}+b\ket{\downarrow_3}\ket{\uparrow_3}),
\end{align}
where $|a|^2+|b|^2=1$.
Then, we analyze whether entanglement monogamy appears due to gravity.
Entanglement monogamy is the condition wherein the 2nd and 3rd particles are maximally entangled, but the 1st particle cannot be entangled with the 2nd or 3rd particle \cite{Coffman}.
The Hamiltonian is given by Eq.~(\ref{1-3}) with $N=3$ and the evolved state at 
the time $t$ is 
$\ket{\Psi(t)}=e^{-i \frac{H}{\hbar}t}\ket{\Psi(0)}$.
The density matrix of the total system is given by Eq. (\ref{d2}) in Appendix D.
The reduced density matrix is given by
\begin{align}
\label{6-2}
\rho_{12}(t)
&=\text{Tr}_3[\ket{\Psi(t)}\bra{\Psi(t)}]\notag\\
&=\frac{1}{2}
\left(
\begin{array}{cccc}
|a|^2&0&|a|^2e^{-i\frac{\Delta_{12}-\Delta_{13}}{\hbar}t}&0\\
0&|b|^2&0&|b|^2e^{i\frac{\Delta_{12}-\Delta_{13}}{\hbar}t}\\
|a|^2e^{i\frac{\Delta_{12}-\Delta_{13}}{\hbar}t}&0&|a|^2&0\\
0&|b|^2e^{-i\frac{\Delta_{12}-\Delta_{13}}{\hbar}t}&0&|b|^2
\end{array}
\right).
\end{align}
We find that all the eigenvalues of the partially transposed matrix 
are positive; hence, there is no entanglement between the 1st and 2nd particles. Similarly, there is no entanglement between the 1st 
and 3rd particles.
This clearly shows the appearance of monogamy in the entanglement
through gravitational interaction.
Interestingly, this property appears even when the 2nd and 
3rd particles are not maximally entangled.

In contrast, the 1st particle is entangled with the composite system of the 2nd and 3rd particles.
We partially transpose the density matrix in Eq. (\ref{d2}) in Appendix D concerning the 1st particle to analyze the entanglement between the 1st particle and the system comprising the 2nd and 3rd particles. Then, we obtain the eight eigenvalues of the partially transposed matrix, from which the negativity is determined as
\begin{align}
\label{6-3}
\mathcal{N}_{1-(2,3)}
&=|a||b|\left|\sin\left[\frac{\Delta_{12}-\Delta_{13}}{\hbar}t\right]\right|.
\end{align}
This negativity has the same form as that of the two-particle system in Eq. (\ref{2-6}) where $a=b=1/\sqrt{2}$. Thus, 
the entanglement 
between the 1st particle and the composite system of the 2nd and 3rd particles
can be regarded as 
the entanglement for the case where $N=2$. 
This results from the initial reduction of the underlying states.
When we select the initial state in Eq. (\ref{6-1}), the underlying basis states of the 2nd and 3rd particles are $\ket{\uparrow_2}\ket{\downarrow_3}$ and 
$\ket{\downarrow_2}\ket{\uparrow_3}$, respectively.
In contrast, when we select the initial state in Eq. (\ref{3-1}), 
the underlying basis states of the 2nd and 3rd particles are $\{\ket{\uparrow_2}\ket{\uparrow_3},~\ket{\uparrow_2}\ket{\downarrow_3},~\ket{\downarrow_2}\ket{\uparrow_3},~\ket{\downarrow_2}\ket{\downarrow_3}\}$, respectively.

Additionally, we consider the partial trace of the density matrix of the total system 
$\ket{\Psi(t)}\bra{\Psi(t)}$, with the initial condition of Eq. (\ref{6-1}), with 
respect to the 1st particle to focus on the entanglement between the 2nd 
and 3rd particles. 
Then, we determine the following negativity of the partially transposed matrix:  
\begin{align}
\label{6-4}
\mathcal{N}_{2-3}
&=|a||b|\left|\cos\left[\frac{\Delta_{12}-\Delta_{13}}{\hbar}t\right]\right|.
\end{align}
From the above results, Eqs. (\ref{6-3}) and (\ref{6-4}), it can be seen that as $\mathcal{N}_{1-2,3}$ increases, $\mathcal{N}_{2-3}$ decreases and vice versa. These properties can be considered as the effects of the entanglement monogamy.

Let us consider the case of $\Delta_{12}=\Delta_{13}$,
wherein the gravitational force between the 1st and 2nd particles is the same as that between the 1st and 3rd particles.
Then, from Eqs. (\ref{6-3}) and (\ref{6-4}), we have that 
\begin{align}
    \label{6-5}
    \mathcal{N}_{1-2,3}=0,\:\mathcal{N}_{2-3}=|a||b|.
\end{align}
In other words, the 1st particle is never entangled with the system of the 2nd and 3rd particles, 
and the 2nd particle is always entangled with the 3rd particle.
In such a situation, the gravitational interaction is defined as
\begin{align}
H_{12}
&=-\frac{\Delta_{12}}{2}\sigma^{(1)}_{z}\otimes\sigma^{(2)}_{z}\otimes I_{3}\notag\\
H_{13}
&=-\frac{\Delta_{12}}{2}\sigma^{(1)}_{z}\otimes I_{2}\otimes\sigma^{(3)}_{z}\notag.
\end{align}
Then, the system of the 1st and 2nd particles evolves in the same
manner as that of the system of the 1st and 3rd particles.
Therefore, the 1st particle is not entangled with the 2nd and 3rd particles.

\section{SUMMARY AND CONCLUSION}

We investigated the quantum many-body effect in the entanglement between a multi-particle system due to gravity.
Our model is the simplest 
extension of Ref.~\cite{NguyenBernards20}. This simplicity allows us to analyze the system exactly. 
We found that a specific particle pair in the three-particle system produces
periodic entanglement. For an $N(>3)$-particle system, quantum entanglement in a specific particle pair may be present initially; however, this 
entanglement tends to disappear through  
entanglement with other particles due to gravity, which plays 
the role of the environment. 
This can be regarded as a decoherence phenomenon due to gravity. 
We estimated the characteristic time of this type of 
decoherence for the first time.
In the case of a two-dimensional system, we derived the state of the system using an analysis similar to that used in the case of the one-dimensional system.
The decoherence in the case of the two-dimensional system was more influential because of the increase in the number of particles.
Furthermore, we discussed the monogamy of the entanglement by considering
a system wherein two out of the three particles in the system were prepared in an initially
entangled state. 
These phenomena of quantum many-body systems are expected to be useful in testing 
the quantum nature of gravity. 

\begin{acknowledgments}
We thank S. Kanno, J. Soda, Y. Nambu, N. Matsumoto, Y. Kamiya,  and H. Suzuki for their insightful. 
communications related to the topic of the present paper. 
This work was supported by the Ministry of Education, Culture, Sports, Science and 
Technology (MEXT)/Japan Society for the Promotion of Science 
(JSPS) KAKENHI Grant  No. 17K05444 (KY).
\end{acknowledgments}

\appendix
\section{GRAVITATIONAL INTERACTION BETWEEN TWO PARTICLES}

The Hamiltonian representing the gravitational 
potential between the $i$-th and $j$-th particles can be defined as
\begin{align}
\label{7-1}
V_{ij}
&=-Gm_im_j
\left(
\begin{array}{cccc}
1/d|j-i|&0&0&0\\
0&1/\sqrt{d^2|j-i|^2+L^2}&0&0\\
0&0&1/\sqrt{d^2|j-i|^2+L^2}&0\\
0&0&0&1/d|j-i|
\end{array}
\right)\notag\\
&=-\frac{Gm_im_j}{2}\left(\frac{1}{d|j-i|}-\frac{1}{\sqrt{d^2|j-i|^2+L^2}}\right)
\left(
\begin{array}{cccc}
1&0&0&0\\
0&-1&0&0\\
0&0&-1&0\\
0&0&0&1
\end{array}
\right)\notag\\
&-
\frac{Gm_im_j}{2}\left(\frac{1}{d|j-i|}+\frac{1}{\sqrt{d^2|j-i|^2+L^2}}\right)
\left(
\begin{array}{cccc}
1&0&0&0\\
0&1&0&0\\
0&0&1&0\\
0&0&0&1
\end{array}
\right)\notag.
\end{align}
Here, the order of the basis is $\{\ket{\uparrow_i}\ket{\uparrow_j},~\ket{\uparrow_i}\ket{\downarrow_j},~\ket{\downarrow_i}\ket{\uparrow_j},~\ket{\downarrow_i}\ket{\downarrow_j}\}$.
Using the Pauli matrices of the individual two-level systems $\sigma^{(i)}_z$ 
and $\sigma^{(j)}_z$, and the unit matrix, the Hamiltonian 
$V_{ij}$ can be defined as 
\begin{eqnarray}
\label{abb}
V_{ij}
=-\frac{\Delta_{ij}}{2}\sigma^{(i)}_z\otimes \sigma^{(j)}_z-\frac{\Delta_{Cij}}{2}I_i\otimes I_j,
\end{eqnarray}
where $\Delta_{ij}$ and $\Delta_{Cij}$ are
\begin{align}
\Delta_{ij}
&=Gm_im_j\left(\frac{1}{d|j-i|}-\frac{1}{\sqrt{d^2|j-i|^2+L^2}}\right)\\
\Delta_{Cij}
&=Gm_im_j\left(
\frac{1}{d|j-i|}+\frac{1}{\sqrt{d^2|j-i|^2+L^2}}\right).
\end{align}
Hence, the total Hamiltonian of the $N$-particle system is
\begin{align}
\label{ab3}
H'
&=-\sum^N_{i<j}\frac{\Delta_{ij}}{2}I_{1}\otimes\cdots\otimes\sigma_{z}^{(i)}\otimes\cdots\otimes\sigma_{z}^{(j)}\otimes\cdots\otimes I_{N}-\sum^N_{i<j}\frac{\Delta_{Cij}}{2}I_{1}\otimes\cdots\otimes\cdots\otimes I_{N}
\nonumber \\
&
=H-\sum^N_{i<j}\frac{\Delta_{Cij}}{2}I_{1}\otimes\cdots\otimes\cdots\otimes I_{N},
\end{align}
where $H$ is defined by Eq.\eqref{1-3}.
For the unitary evolution given by the Hamiltonian $H'$, the second term of Eq.\eqref{ab3} does not contribute because 
$\rho(t)=e^{-iH't/\hbar}\rho(0)e^{iH't/\hbar}=e^{-iHt/\hbar}\rho(0)e^{iHt/\hbar}$.
Hence, it is sufficient to consider the first term in Eq.\eqref{ab3} for our analysis.

\section{DERIVATION OF DENSITY MATRIX FOR THE SYSTEM OF $N$ PARTICLES}

The density matrix for the $N$-particle system is given by
\begin{align}
    \rho(t)
    &=\ket{\Psi(t)}\bra{\Psi(t)}
    =e^{-iHt/\hbar}\ket{\Psi(0)}\bra{\Psi(0)}e^{iHt/\hbar}
\end{align}
where the initial state $\ket{\Psi(0)}$ and Hamiltonian $H$ are given by Eqs. (\ref{1-1}) and (\ref{1-3}), respectively.
We focus on the system comprising the 1st and 2nd particles by tracing over the
Hilbert space of the other particles. Then, the reduced density matrix is 
\begin{align}
    \rho_{12}(t)
    &={\rm Tr}_{3,\cdots,N}[\rho(t)].
\end{align}
By using the Bloch representation, the density matrix is given by
\begin{align}
    \rho_{12}(t)
    &=\frac{1}{4}\sum_{i,j}\lambda_{ij}\sigma_{i}^{(1)}\otimes\sigma_{j}^{(2)},
    \quad
    (i,j=0,1,2,3)
    \label{B3}
\end{align}
where $\sigma_{0}=I$ and the others are Pauli matrices.
Determining the coefficients $\lambda_{ij}$ gives the expression
for the reduced density matrix $\rho_{12}(t)$.
By using ${\rm Tr}[\sigma_{i}\sigma_{j}]=2\delta_{ij}$, the coefficients $\lambda_{ij}$ are obtained as
\begin{align}
    \lambda_{ij}
    &={\rm Tr}_{1,2}[\sigma_{i}^{(1)}\otimes\sigma_{j}^{(2)}\rho_{12}(t)]\notag\\
    &={\rm Tr}_{1,2}[\sigma_{i}^{(1)}\otimes\sigma_{j}^{(2)}Tr_{3,\cdots,N}[\rho(t)]]\notag\\
    &={\rm Tr}_{1,\cdots,N}[\sigma_{i}^{(1)}\otimes\sigma_{j}^{(2)}[\rho(t)]]\notag\\
    &=\bra{\Psi(t)}\sigma_{i}^{(1)}\otimes\sigma_{j}^{(2)}\ket{\Psi(t)}.
\end{align}
After performing complex calculations, we obtain $\lambda_{ij}=0$ except for
\begin{align}
    \lambda_{00}&=1,\notag\\
    \lambda_{01}&=\cos(\frac{\Delta_{12}t}{\hbar})\prod_{i=3}^{N}\cos(\frac{\Delta_{2i}t}{\hbar}),\notag\\
    \lambda_{10}&=\cos(\frac{\Delta_{12}t}{\hbar})\prod_{i=3}^{N}\cos(\frac{\Delta_{1i}t}{\hbar}),\notag\\
    \lambda_{11}&=\frac{1}{2}\Bigl[\prod_{i=3}^{N}\cos(\frac{\Delta_{1i}+\Delta_{2i}}{\hbar}t)+\prod_{i=3}^{N}\cos(\frac{\Delta_{1i}-\Delta_{2i}}{\hbar}t)\Bigr],\notag\\
    \lambda_{22}&=-\frac{1}{2}\Bigl[\prod_{i=3}^{N}\cos(\frac{\Delta_{1i}+\Delta_{2i}}{\hbar}t)-\prod_{i=3}^{N}\cos(\frac{\Delta_{1i}-\Delta_{2i}}{\hbar}t)\Bigr],\notag\\
    \lambda_{23}&=-\sin(\frac{\Delta_{12}t}{\hbar})\prod_{i=3}^{N}\cos(\frac{\Delta_{1i}t}{\hbar}),\notag\\
    \lambda_{32}&=-\sin(\frac{\Delta_{12}t}{\hbar})\prod_{i=3}^{N}\cos(\frac{\Delta_{2i}t}{\hbar}),\notag.
\end{align}

Finally, we derive the density matrix, Eq. (\ref{1-7}), by substituting these coefficients into Eq. (\ref{B3}).

\section{EIGENVALUES OF PARTIALLY TRANSPOSED MATRIX FOR THE SYSTEM OF THREE PARTICLES}

We find that the four eigenvalues of the partially transposed matrix of Eq. (\ref{3-10}) can be defined as
\begin{align}
\label{lambdapm}
\lambda_{\pm}
=\frac{1}{4}\Big(1&\pm \cos(\frac{\Delta_{13}}{\hbar}t)\cos(\frac{\Delta_{23}}{\hbar}t)\notag\\
&-\sqrt{1+\cos^2(\frac{\Delta_{13}}{\hbar}t)\cos^2(\frac{\Delta_{23}}{\hbar}t)\pm2\cos(\frac{2\Delta_{12}}{\hbar}t)\cos(\frac{\Delta_{13}}{\hbar}t)\cos(\frac{\Delta_{23}}{\hbar}t)}\Big),\\
\label{lambdapm'}
\lambda_{\pm}'
=\frac{1}{4}\Big(1&\pm \cos(\frac{\Delta_{13}}{\hbar}t)\cos(\frac{\Delta_{23}}{\hbar}t)\notag\\
&+\sqrt{1+\cos^2(\frac{\Delta_{13}}{\hbar}t)\cos^2(\frac{\Delta_{23}}{\hbar}t)\pm2\cos(\frac{2\Delta_{12}}{\hbar}t)\cos(\frac{\Delta_{13}}{\hbar}t)\cos(\frac{\Delta_{23}}{\hbar}t)}\Big)
\end{align}
where the $\lambda_{\pm}'$s are always positive or zero. Hence, we consider $\lambda_{\pm}$.
When we assume $\lambda_{+}\ge0$, the following inequality holds:
\begin{align}
\label{7-5}
&\left(1-\cos(\frac{2\Delta_{12}}{\hbar}t)\right)\cos(\frac{\Delta_{13}}{\hbar}t)\cos(\frac{\Delta_{23}}{\hbar}t)
\ge0. 
\end{align}
Because
$1-\cos(\frac{2\Delta_{12}}{\hbar}t)\ge0$, we have $\cos(\frac{\Delta_{13}}{\hbar}t)\cos(\frac{\Delta_{23}}{\hbar}t)\ge0$. 
Then, we obtain the inequality as
\begin{align}
\label{7-6}
(1&-\cos(\frac{\Delta_{13}}{\hbar}t)\cos(\frac{\Delta_{23}}{\hbar}t))^2\notag\\
&-(1+\cos^2(\frac{\Delta_{13}}{\hbar}t)\cos^2(\frac{\Delta_{23}}{\hbar}t)-2\cos(\frac{2\Delta_{12}}{\hbar}t)\cos(\frac{\Delta_{13}}{\hbar}t)\cos(\frac{\Delta_{23}}{\hbar}t))\notag\\
&=-2(1-\cos(\frac{2\Delta_{12}}{\hbar}t))\cos(\frac{\Delta_{13}}{\hbar}t)\cos(\frac{\Delta_{23}}{\hbar}t))\notag\\
&\le 0,
\end{align}
where the last inequality holds by $\cos(\frac{\Delta_{13}}{\hbar}t)\cos(\frac{\Delta_{23}}{\hbar}t)\ge0$. Thus, we have 
\begin{align}
\label{7-7}
1-\cos(\frac{\Delta_{13}}{\hbar}t)\cos(\frac{\Delta_{23}}{\hbar}t)
&\le
\sqrt{1+\cos^2(\frac{\Delta_{13}}{\hbar}t)\cos^2(\frac{\Delta_{23}}{\hbar}t)-2\cos(2\frac{\Delta_{12}}{\hbar}t)\cos(\frac{\Delta_{13}}{\hbar}t)\cos(\frac{\Delta_{23}}{\hbar}t)},\notag
\end{align}
which implies that  $\lambda_{-}\le0$. Similarly, when we assume $\lambda_-\ge0$, we have $\cos(\frac{\Delta_{13}}{\hbar}t)\cos(\frac{\Delta_{23}}{\hbar}t)\le0$,
which leads to  $\lambda_{+}\le0$. Therefore, one of the eigenvalues $\lambda_\pm$ necessarily assumes negative values; thus, the negativity is given by Eq. (\ref{3-11}).

\section{DENSITY MATRIX OF THE INITIALLY ENTANGLED SYSTEM}

We consider the three-particle system wherein two particles are initially entangled.
The density matrix is given by
\begin{align}
    \rho(t)
    =e^{-iHt/\hbar}\ket{\Psi(0)}\bra{\Psi(0)}e^{iHt/\hbar},
\end{align}
where the initial state and Hamiltonian for $N=3$ are given by Eqs. (\ref{6-1}) and (\ref{1-3}), respectively.
By evaluating this expression, we eventually obtain
\begin{align}
    \label{d2}
    \rho(t)
    =\frac{1}{2}
    \left(
     \begin{array}{cccccccc}
        0&0&0&0&0&0&0&0\\
        0&|a|^2&a^*be^{-i(\Delta_{12}-\Delta_{13})t/\hbar}&0&0&|a|^2e^{-i(\Delta_{12}-\Delta_{13})t/\hbar}&a^*b&0\\
        0&ab^*e^{i(\Delta_{12}-\Delta_{13})t/\hbar}&|b|^2&0&0&ab^*&|a|^2e^{i(\Delta_{12}-\Delta_{13})t/\hbar}&0\\
        0&0&0&0&0&0&0&0\\
        0&0&0&0&0&0&0&0\\
        0&|a|^2e^{i(\Delta_{12}-\Delta_{13})t/\hbar}&a^*b&0&0&|a|^2&a^*be^{i(\Delta_{12}-\Delta_{13})t/\hbar}&0\\
        0&ab^*&|b|^2e^{-i(\Delta_{12}-\Delta_{13})t/\hbar}&0&0&ab^*e^{-i(\Delta_{12}-\Delta_{13})t/\hbar}&|b|^2&0\\
        0&0&0&0&0&0&0&0
\end{array}
\right),
\end{align}
where the order of the basis is
$\{\ket{\uparrow_1\uparrow_2\uparrow_3}$, $\ket{\uparrow_1\uparrow_2\downarrow_3}$, $\ket{\uparrow_1\downarrow_2\uparrow_3}$, $\ket{\uparrow_1\downarrow_2\downarrow_3}$, $\ket{\downarrow_1\uparrow_2\uparrow_3}$, $\ket{\downarrow_1\uparrow_2\downarrow_3}$, $\ket{\downarrow_1\downarrow_2\uparrow_3}$, $\ket{\downarrow_1\downarrow_2\downarrow_3}\}$.

\section{PPT criterion and negativity}
We consider a system composed of two subsystems A and B.
If the system is separable, the density matrix of the system is given by
\begin{equation}
\rho=\sum_{j}p_{j}\rho_{Aj}\otimes\rho_{Bj},
\end{equation}
where $\rho_{Aj}$ and $\rho_{Bj}$ are the density matrices of the two subsystems and $p_{j}$ is a positive value satisfying
\begin{equation}
\sum_{j}p_{j}=1.
\end{equation}
The density matrix $\rho$ and the density matrix of the two subsystems $\rho_{Aj}$ and $\rho_{Bj}$ are positive definite.
Here, the partially transposed state with respect to subsystem A is
\begin{equation}
\rho^{T_{A}}=\sum_{j}p_{j}(\rho_{Aj})^{T}\otimes\rho_{Bj}.
\end{equation}
Therefore, if the system is separable, the partially transposed state is positive definite because $\rho_{Aj}$ is positive definite.
Hence, if the partially transposed state is negative definite, the system is entangled.
This is known as the positive partial transpose (PPT) criterion, and we introduce negativity to determine whether the state is entangled.
The negativity is defined as
\begin{equation}
\mathcal{N}\equiv\sum_{\lambda_{i}<0}|\lambda_{i}|,
\end{equation}
where $\lambda_{i}$ is the eigenvalue of the partial transposition of the density matrix.
The system is entangled if the negativity is positive.

\end{document}